\documentclass{aa}
\usepackage{graphicx}
\begin{document}
\title{Discovery of New Milky Way Star Clusters Candidates in
the 2MASS Point Source Catalog}
\subtitle{}

    \author{V.D.~Ivanov\inst{1}
      \and
	J.~Borissova\inst{2}
      \and
	P.~Pessev\inst{3}
      \and
	G.R.~Ivanov\inst{3}
      \and
	R.~Kurtev\inst{3}
      }

    \offprints{V.~D.~Ivanov}

    \institute{European Southern Observatory, Ave. Alonso de
    	Cordova 3107, Casilla 19, Santiago 19001, Chile\\
	\email{vivanov@eso.org}
      \and
        Institute of Astronomy, Bulgarian Academy of Sciences, and
        Isaac Newton Institute of Chile, Bulgarian Branch,
        72~Tsarigradsko Chauss\`ee, BG\,--\,1784 Sofia, Bulgaria\\
        \email{jura@haemimont.bg}
      \and
	Department of Astronomy, Sofia University, Bulgaria, and
        Isaac Newton Institute of Chile, Bulgarian Branch,
        5~James Bourchier, 1164 Sofia, Bulgaria\\
	\email{pessev, givanov, rkurtev@phys.uni-sofia.bg}
      }

   \date{Received .. ... 2002; accepted .. ... 2002}

   \authorrunning{Ivanov et al.}
   \titlerunning{Discovery of New Milky Way Star Clusters}

\abstract{A systematic search of the 2MASS point source
catalog, covering 47\% of the sky, was carried out aiming to
reveal any hidden globular clusters in our Galaxy. Eight new
star clusters were discovered by a search algorithm based on
finding peaks in the apparent stellar surface density, and a
visual inspection of their vicinities yielded additional two.
They all are concentrated toward the Galactic plane and are
hidden behind up to ${\rm A}_V$=20 mag which accounts for
their late discovery.
The majority of new clusters are associated with H{\sc ii}
regions or unidentified {\it IRAS} sources suggesting that
they are young, probably similar to Arches or open clusters.
Only one candidate has morphology similar to a globular
cluster and the verification of its nature will require deeper
observations with higher angular resolution than the 2MASS data.
}

\maketitle

\section{Introduction}

There are about 150 known Galactic globular clusters (GC hereafter;
Harris \cite{har96}). The majority of them were discovered through
optical searches, biased against highly obscured objects. Since the
Galaxy is estimated to have 160$\pm20$ GCs (Harris \cite{har91}), a
certain number of GCs may still be hidden behind the Galactic disk.

The Two Micron All Sky Survey (2MASS) offers an opportunity to
carry out a systematic and unbiased search for missing GCs because
it covers in an uniform way the Galactic plane in near infrared
wavelengths $(J, H$ and $K_S$ bands) where the extinction is almost
ten times smaller in comparison with the optical part of the
spectrum (Bessell \& Brett \cite{bes88}; used throughout this
letter).
Using the 2MASS data base Hurt et al. (\cite{hur00}) found two new
GCs: 2MASS GC01 and 2MASS GC02 (see also Ivanov et al.
\cite{iva00}). Later Dutra \& Bica (\cite{dut00}, \cite{dut01})
presented a sample of about 90 new infrared star clusters, stellar
groups and candidates, mostly discovered from visual inspection of
the 2MASS images. Recently, Reyl\'{e} \& Robin (\cite{rey02})
reported two new clusters discovered with DENIS. They applied a
combined surface density--integrated flux--color criterion, that
detected in addition 22 known clusters.

We report the first results of a systematic and objective search
of new clusters in the currently released part of the 2MASS
point--source catalog, covering 47\% of the sky. We also give a
short description of the technique, used to locate cluster
candidates. The list of objects presented here is not aimed to be
complete in any sense.

\section{Search Algorithm}

A simple and robust method, based on the apparent stellar surface
density was chosen to search for obscured clusters. The first
step was to divide the 2MASS point source catalog into spatial
bins. We used square bins, to minimize the computational demands.
Their size was a free parameter, allowing to search for structures
of various scales on the sky. For each bin we stored the total
number of stars, the $K_S$-band luminosity function, and the
distribution of stars along $J-K_S$ color. Effectively, the
results from the first step are two-dimensional histograms on the
sky.

Next, we searched for peaks in the 2-D histogram of total number
of stars in each bin. The background level and its standard
deviation $\sigma$ were calculated from the average number of
stars in the neighboring bins. Our experiments on fields with
known star clusters indicated that the most effective
cluster-finding strategy was to use a two-step criterion:
(i) $3\sigma$ deviation above the background, and
(ii) an excess of 50 or more stars in the bin above the
background.
It proved to work better than the $3\sigma$ excess limit alone,
probably because the
value of $\sigma$ is often ill-defined, especially in fields
with small stellar density.

The results presented here were obtained with parameters
optimized to search for clusters with large apparent sizes on
the sky since those are most likely to be discovered soon -
either serendipitously or after systematic searches, such as the
findings of Hurt et al. (\cite{hur00}) and Dutra \& Bica
(\cite{dut00}, \cite{dut01}). For comparison, the median value
of the half-mass radius for 141 globular clusters with known
structural parameters from Harris (\cite{har96}) is only 1.0
arcmin. The average is $1.3\pm0.8$ arcmin, with a maximum of
4.18 arcmin for $\omega$ Cen. Although the search for open
clusters are not the main objective of this program, we ensured
that they were also likely to be selected with the chosen
parameters. The average size of Galactic open clusters included
in the catalog of Lynga (\cite{lyn95}) is $3.5\pm2.0$ arcmin,
similar to our bin size. Of course, the lower surface density of
open clusters in comparison with globulars makes them more
challenging targets.

Overall, we are probing a different range of the cluster
parametric space than the previous works. For example, we are
sensitive to clusters with larger angular diameters, compared
with the work of Dutra \& Bica (\cite{dut01}) who found more
compact objects, with median size along the large axis of only
1.8 arcmin. The objects found by Reyl\'{e} \& Robin
(\cite{rey02}) are also smaller than 2 arcmin. Smaller bin
sizes than 5 arcmin will constrain the search to the regime of
compact clusters which are usually only partially resolved by
2MASS due to the large pixel size (1 arcsec), and therefore are
not present in the point source catalog. Visual inspections or
flux criteria are better suited for discovering such objects.

Clusters, larger than 5 arcmin are likely to be missed, given
our choice of the bin size. However, they represent a part of
the parametric space that is not likely to yield {\it new}
candidates because such clusters would probably be relatively
nearby. Therefore, they would suffer lesser extinction, and
would have been easily discovered in optical.

The method was implemented as a set of C-based codes in order
to carry out the process automatically. We plan to explore wider
range of parameters in the future, and to shift the bin centers
by half bin size, improving the completeness of the sample.

\section{Results and Discussion}

\subsection{Cluster Parameters}

The search yielded 247 candidates that satisfied the $3\sigma$
and 50 stars excess criteria described in the previous section.
Of those, 105 were known clusters, present in SIMBAD. Incidentally,
2MASS GC01 was rejected based on insignificant peak $(2.5\sigma),$
while 2MASS GC02 was not present in the released point source
catalog. We inspected visually the 2MASS images of the remaining
candidates, and found two more objects. No obvious objects were
present in 134 cases. The basic data for the new clusters is given
in Table~\ref{TblCandidates}. A mosaic of true color images of
nine clusters is shown in Figure~\ref{Fig_CC01}, constructed from
the 2MASS $JHK_S$ images. It is not surprising that all candidates
are situated close to the Galactic plane. This region suffers from
the highest extinction which makes it easy to hide unknown
clusters. All candidates are at least partially resolved, and many
of them show extended emission, that might indicate ionized gas or
faint population, unreachable with the 2MASS data.

\begin{table}[t]
\begin{center}
\caption{Parameters of the cluster candidates. The first eight
objects were identified by the automatic algorithm, and the
last two were found after a visual inspection. See
Sec.~\ref{IndivObj} for comments on individual objects.}
\label{TblCandidates}
\begin{tabular}{l@{}c@{ }c@{}c@{}c@{}c}
\hline
\multicolumn{1}{c}{ID} &
\multicolumn{1}{c}{R.A.  Dec.} &
\multicolumn{1}{c}{{\it l   b}} &
\multicolumn{1}{c}{D} &
\multicolumn{1}{c}{$K_S$,$J$-$K_S$,} &
\multicolumn{1}{c}{$A_V$} \\
\multicolumn{1}{c}{CC} &
\multicolumn{1}{c}{(J2000.0)} &
\multicolumn{1}{c}{} &
\multicolumn{1}{c}{$\arcmin$} &
\multicolumn{1}{c}{$H$-$K_S$} &
\multicolumn{1}{c}{mag} \\
\multicolumn{1}{c}{} &
\multicolumn{1}{c}{} &
\multicolumn{1}{c}{} &
\multicolumn{1}{c}{} &
\multicolumn{1}{c}{mag} &
\multicolumn{1}{c}{} \\
\hline
01 & 05:13:26 $+$37:27.0 & 169.19 $-$0.90 & 3.0 & 7.5 0.5 0.6 & 6-13 \\
02 & 06:15:53 $+$14:16.0 & 196.21 $-$1.20 & 2.0 & 6.3 1.2 1.0 & 6-13 \\
03 & 06:59:14 $-$03:55.0 & 217.30 $-$0.05 & 2.8 & 6.2 0.8 0.1 & 6-13 \\
04 & 07:00:32 $-$08:52.0 & 221.85 $-$2.03 & 4.0 & 7.5 1.8 0.9 & 9-17 \\
05 & 07:00:51 $-$08:56.5 & 221.96 $-$1.99 & 2.4 & 6.8 0.3 0.0 & 9-17 \\
06 & 07:24:14 $-$24:38.0 & 238.48 $-$4.28 & 4.5 & 6.5 0.8 0.7 & 9-17 \\
07 & 07:30:40 $-$15:18.0 & 230.98 $+$1.49 & 2.8 & 6.1 0.7 0.6 & 9-17 \\
08 & 08:19:10 $-$35:39.0 & 254.01 $+$0.25 & 2.8 & 6.3 0.5 0.7 & 4-12 \\
\hline
09 & 06:59:43 $-$04:04.0 & 217.49 $-$0.02 & 1.0 & ...         & 4-12 \\
10 & 08:18:28 $-$35:47.5 & 254.05 $+$0.05 & 0.5 & 9.7 2.0 0.8 & 12-20 \\
 & & & & 10.3 2.4 1.0 & \\
\hline
\end{tabular}
\end{center}
\end{table}

\begin{figure}
% \resizebox{\hsize}{!}{\includegraphics{rgb.ps}}
\caption{Mosaic of true color images for nine of the new
clusters listed in Table~\ref{TblCandidates}.
The top left is CC\,01, the numbers increase from left to right, 
and toward the lower rows. CC\,09 is skipped. Blue is $J$, green 
is $H$, and red is $K_S$. North is up and East is to the left. 
The individual image are 4.8$\times$4.8 arcmin.}
\label{Fig_CC01}
\end{figure}

The cluster coordinates are impossible to estimate by fitting of
radial profiles because of the relatively low number of members.
This forced us to apply an alternative method for finding the
centers. We adopted a trial center, and minimized the sum of the
distances to the point sources within 1.5 arcmin. Whenever
possible, obvious non-members were excluded based on the
color-magnitude diagrams. Then we moved the trial center in a
rectangular grid pattern across the face of the cluster. The
coordinates, presented here are accurate within 30 arcsec, and
agree within this uncertainty with visual estimates. In both
cases we used 2MASS world coordinates, from the point source
catalog, or from the image header, respectively. The diameters
were determined after visual inspection of the K-band images.
They should be considered lower limits because some fainter stars
may well be bellow the detection limit of the 2MASS atlas. The
total magnitudes are obtained with aperture measurements, with
diameters 100-180 arcsec. To remove the contribution from the
foreground stars we subtracted the flux measured through the same
aperture near the clusters from the flux measured at the cluster
position. The large variations of the extinction and the apparent
stellar density lead to errors of about 0.5 mag.

We carried out a thorough search for known objects near the new
clusters. The majority of them were found to be associated with
H{\sc ii} ionized regions indicating that they may be young
clusters, perhaps similar to the recently discovered Arches
cluster. The SIMBAD identifications are discussed in
Sec.~\ref{IndivObj}. The most promising candidate for an unknown
globular cluster is CC01, based on its appearance. CC08 and CC10
possess morphologies typical of open clusters.

\begin{figure}
\resizebox{\hsize}{!}{\includegraphics{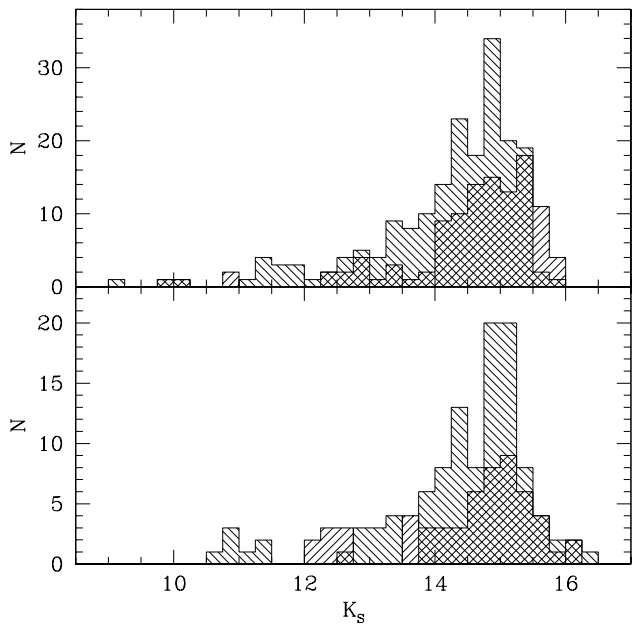}}
\caption{Example of luminosity functions of cluster stars.
The top panel is CC01. The histogram, shaded top left to bottom
right refers to the stars within 3 arcmin from the cluster center.
The histogram, shaded from top right to lower left includes stars
from a circular annulus at about 10 arcmin from the cluster center.
Both regions have the same area. The bottom panel is for CC08, the
central radius is 1.5 arcmin, and the annulus is at about 5 arcmin.}
\label{FigHist}
\end{figure}

\begin{figure}
\resizebox{\hsize}{!}{\includegraphics{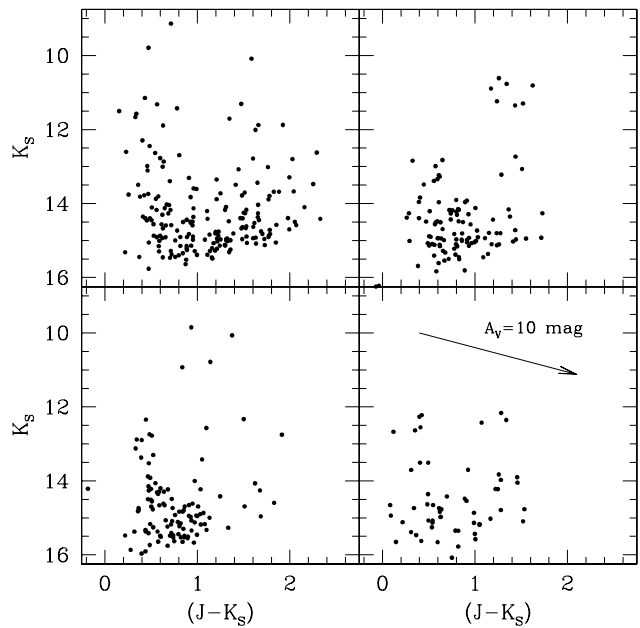}}
\caption{Color-magnitude diagrams for CC01 (left) and CC08 (right).
The top panels show all stars within 3 and 1.5 arcmin from the
cluster centers, respectively. The lower panels show stars in
annuli with the same areas as the central regions, at about 10 and
5 arcmin from the cluster centers, respectively. The excess of
red cluster members is evident in both cases. A reddening vector,
corresponding to visual extinction $\rm A_V$ = 10 mag is also
shown.}
\label{FigCMD}
\end{figure}

To verify further the nature of our candidates we constructed
luminosity functions (LF hereafter) of the areas near the stellar
surface density peaks, and compared them with LFs of circular
regions with the same areas, well away from the objects. An
example is shown in Figure~\ref{FigHist}. The excesses of stars
at the alleged cluster positions are obvious despite of the low
statistics of individual bins. The color-magnitude (CMD hereafter)
and color-color diagrams provide an additional test of our
candidates. Figure~\ref{FigCMD} shows the CMDs of CC01 and CC04.
An excess of stars is evident in both cases. However, these
diagrams are of little help for unobscured or lightly obscured
clusters.

The estimate of the extinction toward the clusters is a problem
on its own. It is impossible to determine whether we see the tip
of the main sequence or the the tip of the red giant or red
supergiant branch, given the relatively shallow depth of the
2MASS point source photometry. We adopted a simple foreground
screen, although it is clear that some of the clusters suffer
from differential reddening. Next, we made two extinction
estimates, assuming that the reddest sequence is either at
$J-K_S\sim0$ mag, typical for the blue stars, or at 0.9-1 mag,
which is the usual color of the red supergiant and red giant
branches. We neglected completely any metallicity effects because
they are much smaller than the uncertainties in the $E(J-K_S)$
which are of order of 0.5 mag, leading to errors of $\sim4$ mag
in $A_V$. The results are included in
Table~\ref{TblCandidates}. Using the
stellar colors to estimate the extinction toward clusters inside
our Galaxy is more reliable than far-infrared techniques (i.e.
Dutra \& Bica \cite{dut00}), because the background dust
emission can easily be confused for emission from foreground
obscuring material.

\subsection{Comments on Individual Objects\label{IndivObj}}

{\bf - CC01} A rich cluster, well resolved by the 2MASS. It
resembles morphologically a globular or a compact young cluster,
similar to Arches. Further observations are needed to solve
this ambiguity. The location of the candidate coincides with an
emission nebula Min\,2-58 (Minkowski \cite{min48}), clearly seen
on both blue and red DSS images (while the stellar cluster is
not). A radio-selected H{\sc ii} region within 30 arcsec was
reported by Lockman (\cite{loc89}).\\
{\bf - CC02} Resolved by the 2MASS. Extended nebulocity is
present on both blue and red DSS images. There is an indication
for faint stellar population on the red image. An H{\sc ii}
region SH\,2-269\,B (Sharpless \cite{sha59}) lies within the
borders of our candidate. Water maser emission was detected by
Comoretto et al. (\cite{com90}) indicating that this is a young
cluster embedded in a dense molecular core.\\
{\bf - CC03} Partially resolved by the 2MASS. Compact (D$\sim 1$
arcmin) nebulocity is visible on the red DSS image, with no
trace of the embedded stars. An H{\sc ii} region BFS\,56,
associated with a molecular cloud was found at the same position
by Blitz, Fich, \& Stark (\cite{bli82}). A nearby IRAS source
(06567-0350) was identified by Magnier et al. (\cite{mag99}) as
possible young stellar object. Therefore, the found cluster is
likely young.\\
{\bf - CC04} Well resolved on the 2MASS images. Extended
nebulocity is visible on the red DSS plate, with a few bright
stars that might be associated with it. Blitz et al.
(\cite{bli82}) reported an H{\sc ii} region BFS\,64 nearby.\\
{\bf - CC05} Partially resolved by the 2MASS. A compact nebulocity
(D$\sim 1$ arcmin) is visible on the red DSS image, with no
associated stars. Fich \& Terebey (\cite{fic96}) reported a cloud
(221.9-2.0B) that might hide on-going star formation. Indeed, a
young stellar object CPM\,33 was discovered by Campbell, Persson,
\& Matthews (\cite{cam89}), which leads as to believe that this
may be a young cluster.\\
{\bf - CC06} Partially resolved by the 2MASS. A patchy nebula
with a few stars near the center is visible on the red DSS image.
However, their association is not obvious due to heavy foreground
contamination. The nebula is included in the H{\sc ii} region
catalog of Brand, Blitz, \& Wouterloot (\cite{bra86}) as
BRAN\,22C. CO emission, associated with this object was detected
by Brand et al. (\cite{bra87}), suggesting the presence of a
dense core.\\
{\bf - CC07} Well resolved by the 2MASS. A patchy faint nebula
with a few stars near the center is visible on the red DSS image.
Some of the stars coincide with the brighter patches, indicating
a physical association. The nebula is cataloged by Sharpless
(\cite{sha59}) as H{\sc ii} region SH\,2-299\,B.\\
{\bf - CC08} Well resolved on the 2MASS images. It is visible on
the red DSS image as a loose concentration of faint stars,
indicating even lower extinction than the one estimated from the
near-infrared color-magnitude diagram (Figure~\ref{FigCMD}, left).
The morphology suggests that this might be an open cluster.\\
{\bf - CC09} Well resolved by the 2MASS. Visible on the red DSS,
suggesting low extinction. We refrained from measuring the
cluster brightness because of the faint apparent magnitude and the
foreground contamination. Resembles an open cluster or a distant
OB association.\\
{\bf - CC10} Partially resolved by the 2MASS. Compact nebulocity
(D$\sim$1 arcmin) visible on the red DSS image. It is associated
with IRAS\,08165$-$3538, and appears to be a compact young cluster
that does not match our search criteria for richness. It was
observed in the near infrared as an unresolved source by Liseau
et al. (\cite{lis92}) and Lorenzetti et al. (\cite{lor93}). Their
measurements are also listed in Table~\ref{TblCandidates} (last
row), and agree with ours, within the uncertainties.

\section{Summary}

A systematic search throughout the 2MASS point source catalog
yielded ten new stellar clusters. One may be a new globular, two
have morphologies consistent with open clusters, and the rest are
likely compact young clusters. Further analysis and deeper imaging
with higher angular resolution is needed to verify their true
nature.

\begin{acknowledgements}
This publication makes use of data products from the Two Micron
All Sky Survey, which is a joint project of the University of
Massachusetts and the Infrared Processing and Analysis
Center/California Institute of Technology, funded by the National
Aeronautics and Space Administration and the National Science
Foundation. This research has made use of the SIMBAD database,
operated at CDS, Strasbourg, France.
\end{acknowledgements}

\end{document}